\documentclass[twocolumn,a4]{revtex4-1}
\usepackage{graphicx}
\usepackage{epstopdf}
\usepackage{longtable}

\begin{document}

\title{Quantum mechanics: A new chapter?}

\author{Werner A Hofer \\
  Department of Physics, The University of Liverpool, UK}

\begin{abstract}
 We review the conceptual problems in quantum mechanics on a fundamental level. It is shown that the proposed model of extended electrons and a clear understanding of rotations in three dimensional space solve a large part of these problems, in particular the problems related to the ontological status and physical meaning of wavefunctions. It also solves the problem of non-locality. The experimental results obtained in Yves Couder's group and theoretical results by Gerdard Gr\"ossing indicate that the wave-like distribution of trajectories of electrons in interference experiments are most likely due to the quantized interactions leading to a discrete set of transferred momenta. A separate experimental confirmation of this
 interpretation for double-slit interferometry of photons has been given by the group of Steinberg.
\end{abstract}

\maketitle


\section{Introduction}

There exists a peculiar fact, not too frequently noted, about the historical development of quantum mechanics. Although the framework is axiomatic, and although its implications have been under attack nearly from the beginning, very few attempts have been made to refer its axioms to a physical basis. Instead it is frequently claimed that the laws of microphysics are at once superior to common human understanding and correctly described by the fundamental equations of quantum physics.
From an epistemological point of view these two statements seem to contradict each other: because if microphysics can be described consistently by mathematics, why not by an, equally logical, framework of fundamental notions and concepts? To this objection conventional reasoning usually replies that the notions we employ are classical, and the notions of microphysics far from classical indeed. So there is no way to describe our experiences in microphysics by a logically consistent framework of classical concepts.
But even if this were the case, and we are forced to accept the peculiarities of non-locality \cite{EPR,aspect82}, individual particles being in fact ensembles of particles \cite{bohm52}, measurements without any interference with measured objects \cite{kwiat95}, infinities not being infinite \cite{schweber94}, or spreading wave packets not really spreading \cite{heisenberg77}, the question seems legitimate, whether all these effects occur in reality, or rather in some logical representation by a theory, which at once employs and contradicts classical physics.

In the usual course of scientific development there exists an intricate balance between speculative or theoretical concepts and experimental results: and it is usually thought that progress is made by an interplay between theory (or creativity) and measurement. This is, at least, what the textbooks tell us. In scientific revolutions, the situation is a little different: an idea forces the reorientation of most of the - already accumulated - knowledge, and in this case the meaning of the facts may be subject to sudden changes.
The basic situation, today, is again a little different: we do have a theory, which describes, what we measure, it just does not make sense. One could call that a pre-revolutionary situation. There are generally two different categories of approaches to this problem \cite{adler09}:

\begin{itemize}
\item A formal approach: we accept the axioms, and just tinker with their interpretation.
\item A realistic approach: we do not a priori accept the axioms, but think of micro physical objects in terms of physical concepts.
\end{itemize}

Most previous theories used the formal approach.
The problem of the second approach seems to be, that it must be based on physical concepts. And since the only concepts of matter in classical physics are particles or waves, the few existing approaches generally use one or the other, or a combination of both. De Broglie \cite{debroglie64} saw quantum waves as real waves, guiding a singularity in random motion within this wave and which was identified as the particle.
Even in this case the particle, the point-particle used in most of Newtonian physics, remained the primary entity. But considering, that we also know about the wave properties of matter, which is theoretically expressed by de Broglie's relations \cite{debroglie25} and experimentally verified by diffraction experiments \cite{davisson27}, this choice seems far from natural and rather arbitrary.
However, a local and realistic formulation of micro physics, which has to include, at least as its limit of formalization, the existing axioms of quantum theory, is confronted with several related obstacles.

\begin{enumerate}
\item The Uncertainty Relations \cite{heisenberg27}.
\item The intrinsic interactions due to electric charge.
\item The experimental proof against local and realistic theories \cite{aspect82}.
\end{enumerate}

The uncertainty relations, if interpreted as a physical axiom, prohibit any local and realistic interpretation, because their immediate consequence is a spreading of wave packets \cite{debroglie88}: these waves therefore cannot be physical.
If electrons are interpreted as wavelets, the problem of electrostatic interactions must be accounted for. In case no solution is found, the electron in motion cannot be interpreted as a wave type structure, because the inevitable self interactions must lead to acceleration effects bound to destroy the original wave structure. In this case the interpretation is not even self-consistent.
There exist several experimental loopholes to these experiments which restrict their validity. But assuming, that even more precise experiments yield the same result, i.e. a violation of Bell's inequalities \cite{bell64} and thus a contradiction with any local and realistic framework, the problem must be solved from a theoretical, rather than an experimental angle.
We show in this paper, which was presented at the conference in V\"axjo, that all of these problems can be overcome within a theoretical framework composed of extended electrons and formulated in terms of geometric algebra.

\section{The model of extended electrons}

The standard interpretation of quantum mechanics is based on a remarkable statement about the physical properties of electrons: electrons, it states, are point particles. As such, they have no extension, no intrinsic properties, are radially symmetric, and their rotation does not have a moment of inertia. This, we think, is the fundamental problem of all conventional interpretations of quantum mechanics, because it requires very complicated models to reconcile this basic statement with the observable properties of electrons, like their wavelength, or their magnetic moments. However, new experiments, undertaken in the last twenty years, suggest that this statement is actually incorrect.

\subsection{Experimental evidence}

The presentation of a model  of extended electrons can start with experimental evidence. It can be shown, that surface electrons of noble metal silver surfaces, which have a band energy of about 80 meV, show a density distribution of electron charge, which cannot be described as the consequence of a probability distribution of detection events, without violating the Uncertainty relations for the local uncertainty by approximately two orders of magnitude. This result was obtained in the following way (for technical details see the relevant paper \cite{hofer2012a}).

First, it was assumed that electrons from the metal surface tunnel into a single radially symmetric state at the tip of a scanning tunneling microscope (STM, see the following review \cite{hofer2003}). This is a widely used and very successful approximation in condensed matter theory. In this case the measured current in the STM is proportional to the distribution of electron charge density at the surface. Second, it was assumed that the energy uncertainty of these surface state electrons must be smaller than the band energy of 80 meV. This leads to a momentum uncertainty in two dimensions and, consequently, to a local uncertainty of about 350 pm.

The experimental error in STM experiments under cryogenic conditions is about 0.05 pm.
If a feature is about 30pm high, an experimental error of 0.1 pm leads to relative error for a feature measurement $\Delta z/z_0$ of 0.3\%. Under these conditions the standard deviation of a position measurement, and the measurement of the electron density distribution is a position measurement, cannot be be larger than the experimental lateral resolution, which for an STM experiment is about 20 pm. Given that the standard deviation, or the local standard deviation from the Uncertainty relations is about 350 pm, we arrive at a clear contradiction: a feature of this height cannot be measured with this precision, unless the electron's energy is about 1000 eV \cite{hofer2012a}. The message of this analysis is clear: electrons are not point particles, or if they are, then their behavior in these experiments does not comply with the uncertainty relations.

\subsection{Postulates}

On the basis of experimental evidence that electrons are extended particles, a new framework can be based on four distinct postulates, formulated as follows \cite{hofer2011}:

\begin{enumerate}
\item The wave properties of electrons are a {\em real} physical
property of electrons in motion.
\item Electrons in motion possess intrinsic electromagnetic potentials.
\item The magnetic moment of electrons is due to these
electromagnetic potentials.
\item In equilibrium the energy density
throughout the space occupied by a single electron is constant.
\end{enumerate}

The first postulate accounts for the analysis of high-resolution STM experiments. The second postulate then rephrases the energy principle. If the kinetic energy density of moving electrons varies due to their wave properties, then a complementary potential {\it must} exist, which acts as an energy reservoir for the variation in kinetic energy density. Magnetic properties, the electron's spin, can only be related to these potentials, so the potentials must be vector like. The last postulate deals with variable electrostatic potentials from the viewpoint of an extended electron. If this electron encompasses the whole negatively charged shell of a hydrogen atom, then something must prevent its collapse due to the principle of energy minimization. This postulate introduces eigenvalue equations into the problem of electron dynamics, which are the basis of our description of atomic scale systems in Physics, Chemistry, and Biology.

\subsection{Wavefunctions}

The primary variable in condensed matter theory is the density of electron charge. One can avoid dealing with awkward units by transforming the density of electron charge into the number density of electrons, which carries the unit of volume to the power of minus one, or the same unit as the square of the scalar wavefunctions in Schr\"odinger's theory of the hydrogen atom \cite{schrodinger26}. If the direction of motion is parallel to the $z$-axis, then the mass density for an electron of velocity $u$ will be:
\begin{eqnarray}
\rho(z,t) = \frac{\rho_{0}}{2}\left[1 + \cos\left(\frac{4 \pi}{\lambda} z - 4 \pi \nu t\right)\right],
\end{eqnarray}
where $\lambda$ and $\nu$ are in some way related to the velocity of the electron, and $\rho_0$ is the mass-density amplitude. The kinetic energy density $\omega_{kin}$ of this electron is therefore:
\begin{eqnarray}
\omega_{kin} &=& \frac{1}{4} \rho_0 u^2 \left[1 + \cos\left(\frac{4 \pi}{\lambda} z - 4 \pi \nu t\right)\right]\nonumber \\ &=&
\frac{1}{2}\rho_0 u^2\cos^2\left(\frac{2 \pi}{\lambda} z - 2 \pi \nu t\right)
\end{eqnarray}
It is clear that this form of the density violates the energy principle, because the energy density varies with time and also location. A remedy to this problem is the introduction of electromagnetic fields ${\bf E},{\bf H}$, which are transversal with the field components described by:
\begin{eqnarray}
{\bf E} &=& {\bf e}_1 E_0 \cos\left(\frac{2 \pi}{\lambda} z
- 2 \pi \nu t +
\phi\right)  \nonumber \\
{\bf H} &=& {\bf e}_2 H_0 \cos\left(\frac{2 \pi}{\lambda} z
- 2 \pi \nu t + \phi\right)
\end{eqnarray}
Here, ${\bf e}_1$ and ${\bf e}_2$ are unit vectors in $x$ and $y$ direction, respectively, and an additional phase $\phi$ shall account for energy conservation. Note that these fields are different from standard electromagnetic fields, as their propagation velocity is not equal to the speed of light, but equal to the mechanical velocity of the electrons. Qualitatively, they are therefore a new class of electromagnetic entities, which we interpret as the {\em spin} of the electron in the following. The directions of the field vectors define a positive and a negative helicity, which in the model correspond to the two spin directions. The spin then
is either parallel or anti parallel to the direction of motion, and it corresponds to a Poynting-like vector obtained by the geometric product of ${\bf E}$ and ${\bf H}$.  In geometric algebra, a product between two frame vectors carries an imaginary unit, since
${\bf e}_1 {\bf e}_2 = i {\bf e}_3$ \cite{doran2002}.
If $\phi = \pi/2$, the spin is described by:
\begin{equation}
{\bf S} = {\bf E} {\bf H} = i {\bf e}_3 E_0 H_0
\sin^2\left(\frac{2 \pi}{\lambda} z
- 2 \pi \nu t\right)
\end{equation}
The energy contained in the spin component of the electron's energy distribution is, from classical electrodynamics \cite{Jackson}:
\begin{eqnarray}
\omega_{field} &=& \frac{1}{2} \epsilon_0{\bf E}^2 + \frac{1}{2}\mu_0 {\bf H}^2 \nonumber \\
&=& \left(\frac{\epsilon_0 E_0^2}{2} + \frac{\mu_0 H^2}{2}\right)  \sin^2\left(\frac{2 \pi}{\lambda} z - 2 \pi \nu t\right)
\end{eqnarray}
And if we define the amplitude of the spin components by:
\begin{equation}
\frac{\epsilon_0 E_0^2}{2} + \frac{\mu_0 H^2}{2} = \frac{1}{2}\rho_0 u^2,
\end{equation}
then the total energy density of the electron in motion will be equal to the energy density of its inertial mass:
\begin{equation}
\omega_{kin} + \omega_{field} = \frac{1}{2} \rho_0 u^2
\qquad W_{el} = \int_{V} \frac{1}{2} \rho_0 u^2 dV = \frac{1}{2} m u^2
\end{equation}
$m$ is the inertial mass of an electron. It is interesting to note that the spin component of a photon will also be parallel or anti-parallel to the vector of motion. However, since photons do not possess mass components, their algebra in three dimensional space is different; they show a $2\pi$ instead of a $4\pi$ symmetry, which accounts for the fact that they are spin-one rather than spin-half entities. We shall elaborate on the spin of photons in the section about quantum paradoxes.

In standard theory, wavefunctions are
usually described mathematically, e.g. by their algebra or their vector space. This is acceptable, if one {\it does not know}, what they actually are, but becomes rather redundant, if they correspond to aspects of the physical properties of electrons. We consider this aspect, establishing a link between physical variables in three dimensional space and wavefunctions, as the main achievement of the model. All other advances follow from these insights.
It was known previously, that (i) wavefunctions are complex numbers, (ii) that their square corresponds to a number density, (iii) that these complex numbers themselves do not describe spin, (iv) and that they comply with the Schr\"odinger equation. In our model we find that the electron is described by two components, a mass-density component and a field component. We also found that the mass component is scalar, while the spin component is a pseudo-vector (a structure described by the geometric product of two vectors \cite{doran2002}). In the most general case the first component is then described by the square root of the number density, which is equal to the mass density in atomic units. Thus:
\begin{equation}
\psi_m = \rho^{1/2} \qquad \rho = \rho_0 \cos^2\left(\frac{2 \pi}{\lambda} z - 2 \pi \nu t\right)
\end{equation}
Given that we know that wavefunctions are the square root of number densities, the unit has to be the square root of the number density as well. However, the number density in atomic units is also equal to the energy density. Then the spin component can be formalized as the square root of the spin times a unit vector ${\bf e}_3$ times the imaginary unit:
\begin{equation}
\psi_s = i {\bf e}_3 S^{1/2} \qquad S = S_0 \sin^2\left(\frac{2 \pi}{\lambda} z - 2 \pi \nu t\right) \qquad S_0 = \rho_0
\end{equation}
The wavefunction of an electron is then:
\begin{equation}
\psi = \psi_m + \psi_s = \rho^{1/2} + i {\bf e}_3 S^{1/2}
\end{equation}
Complex conjugation within this model means a change of spin from positive to negative, or a transposition of electromagnetic field vectors:
\begin{equation}
\psi^* = \rho^{1/2} - i {\bf e}_3 S^{1/2}
\end{equation}
The wavefunction in this case is a multivector in geometric algebra, with a scalar component and a pseudovector component. Both of these components are even elements of the multivector algebra in three dimensional space.
The square of the wavefunction then simultaneously complies with the Born rule \cite{born26},
is equal to the inertial number density, and hides all intrinsic properties contained in the oscillating density components and field components:
\begin{equation}
\psi^* \psi = \rho + S = \rho_0 = \mbox{constant}
\end{equation}
That wavefunctions are in the most general case multivectors can be seen in many-body physics, where they are necessary to capture spin properties in a solid. However, for single electrons, it is possible to reduce the problem by retaining the spin direction as a hidden variable and describing the wavefunction by a complex number, where the imaginary components are due to the fields. In this case a Schr\"odinger-like wavefunction of a free electron can be described as a plane wave \cite{schrodinger26}:
\begin{equation}
\psi_S = \rho_0^{1/2} \exp\left[i\left(\frac{2 \pi}{\lambda}z - 2 \pi \nu t\right)\right]
\end{equation}
The last problem, how the electron as a physical object with finite extensions can be stable, is solved by an additional cohesive potential, which counteracts the Coulomb repulsion of electron charge and has a magnitude of -8.16eV \cite{hofer2011}.

\subsection{Electron dynamics}

Some of the most interesting effects in atomic scale physics have been impossible to analyze in detail within the conventional framework. For example, it is known that the wavelength of electrons depends on the velocity, but it was not known, how this wavelength changes if the velocity changes. It is clear that the group velocity, on the basis of Planck's and de Broglie's relations, is the same as the mechanical velocity of the electron since:
\begin{equation}
v_g = \frac{d \omega}{d k} = \frac{d (mu^2/2 \hbar)}{d (mu/\hbar)} = u
\end{equation}
If this electron is brought into an external field $\phi$, four distinct processes are bound to happen:
\begin{enumerate}
\item The electron velocity will change.
\item The electron density distribution will change.
\item The electromagnetic field components will change.
\item The intensity of the external field will change.
\end{enumerate}
All four processes are captured in the following equation, which could be called the {\em local Ehrenfest theorem}:
\begin{equation}
{\bf F} = - \nabla \phi = \rho_0 \frac{d {\bf u}}{dt}
\end{equation}
Here, ${\bf F}$ is the force on the density $\rho_0$, $\phi$ the potential, and ${\bf u}$ the velocity of the electron.
The reason that the classical result - a force on the inertial mass of the electron - is recovered, is that the density components and spin components change in an opposite manner:
\begin{equation}
\rho + S = \rho_0 = \mbox{constant} \quad \Rightarrow \quad \dot{S} = - \dot{\rho}
\end{equation}
This insight has consequences in a number of processes. For example, it means that an electromagnetic field, which is adsorbed by an electron, will lead to an increase of the field components, therefore a decrease in the density components and consequently an acceleration of the electron. This has been analyzed in some details in the model for Compton scattering \cite{hofer2011}. It also means that there must be a connection between the frequency of an adsorbed photon, the frequency of the electron, and an external electrostatic field, which accounts for the photoelectric currents at metal surfaces. And finally, it means that there is a connection between external vector potentials  and wavevectors of electrons, which becomes measurable in Aharonov-Bohm effects. The underlying principle is always the same: an interplay between field components and mass components, which is affected by external potentials or dynamic fields.

\section{Quantum paradoxes}

Here, we review some of the paradoxes in the conventional framework and show their resolution within the new theoretical models available today.

\subsection{Wavefunction collapse}

This is one of the most problematic of all notions in the conventional theory, due to to two features: (i) A measurement changes the information about a system. As the wavefunction relates to this information, the wavefunction in the measurement will change. (ii) It is unclear, how this could actually happen, since no physical mechanism to this end is known. The answer to the problem within the present context, and for a spin measurement is the following.
\begin{figure}
\centering
\includegraphics[width=\columnwidth]{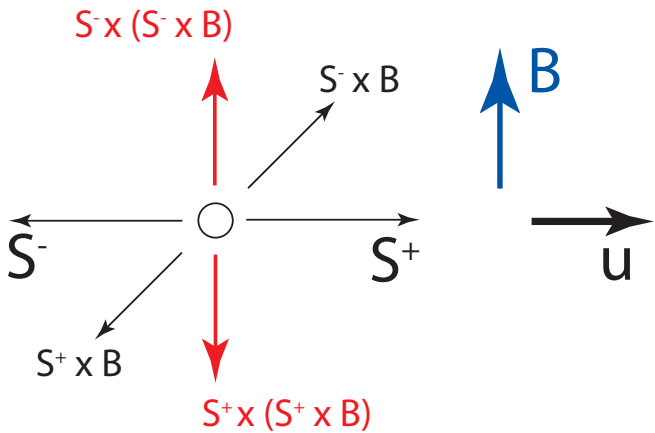}
\caption{{\bf Stern-Gerlach experiments} An electron moving along the $z$-axis with its spin vector parallel ($S^+)$ or anti parallel ($S^-)$ to the vector of motion ${\bf u}$ is subject to a measurement in an external field ${\bf B}$. The spin vector is rotated out of its groundstate position into a direction parallel or anti parallel to the magnetic field vector, which leads to a deflection in a field gradient either in the direction of the field or from the direction of the field.}\label{sg-fig}
\end{figure}
The ensemble of electrons is thought to be composed of an equal number of spin-up (${\bf S}$ parallel to the vector of motion, and spin-down (${\bf S}$ antiparallel to the vector of motion) electrons. Both of these states are isotropic with respect to rotations in a plane perpendicular to the vector of motion. If the electrons are completely free then a magnetic field will alter their trajectory according to the classical Lorentz forces. However, if they are not free, but move along a constrained trajectory, e.g. within the field of a central nucleus, then their spin will be affected by the external field. These changes are described by a modified Landau-Lifshitz equation \cite{hofer2011}:
\begin{equation}
{\bf S} = {\bf e}_S \cdot S \qquad \frac{d {\bf e}_s}{dt} = \mbox{const} \cdot {\bf e}_S \times \left({\bf u} \times \frac{d {\bf B}}{dt}\right)
\end{equation}
This results in a rotation of the ${\bf S}$ vector either parallel or antiparallel to the external magnetic field, which, in a Stern-Gerlach experiment, will lead to a deviation of the trajectory either parallel or antiparallel to the field vector. In this case the wavefunction changes, because the spin component changes (see Fig. \ref{sg-fig}). However, it does not collapse into a definite state - which is usually thought to occur when spin, which is isotropic in three dimensional space, is measured - but it only reveals the direction of the spin vector with respect to the vector of motion. The crucial omission in the conventional framework in this case is the possibility that measurements directly affect the spin properties of a system.

\subsection{Schr\"odinger's cat}

Schr\"odinger's original paper \cite{schrocat} tried to demonstrate the logical problems of the conventional interpretation of quantum mechanics, which held that the wavefunction collapses in an Einstein-Podolksy-Rosen experiment, by a simple thought experiment. A live cat is put into a container with a small radioactive source, a Geiger counter, and a cynanide container, which will be broken once the Geiger counter registers the decay of an atom of the source. If it is assumed that the system can be described by wavefunctions, then a proper description before we open the container is a combination of life and dead cat. In an everyday environment, such a combination does not make sense, because the cat is either alive or dead.
The solution from the viewpoint of the new model is straightforward. If wavefunctions capture physical properties of atomic scale systems, then they must be composed of components related to either the electron density or field components. For a cat, such a wavefunction does not exist, therefore a measurement cannot change this hypothetical wavefunction. Even if it existed, it would be scientifically meaningless since the number of variables in a wavefunction of more than 150 electrons is larger than the number of all particles in the universe: such a wavefunction therefore is not a sensible scientific concept (see Walter Kohn's analysis of this problem, the van Vleck catastrophe, in the Reviews of Modern Physics \cite{kohn99}). Therefore the wavefunction can also not change the state of the cat from alive to dead or vice versa. In colloquial English, one could call this problem a typical red herring. In addition, as shown in the next section, the wavefunction in Einstein-Podolsky-Rosen experiments does not actually collapse.

\subsection{Einstein-Podolsky-Rosen (EPR)}

This paradox and its discussion within the conventional framework contain several misconceptions. The first is that a spin measurement somehow selects spin from isotropic states with respect to rotations to result in a single well defined state. As discussed above, this omits the possibility that spin is changed in a measurement even though it is isotropic. The second is that the framework of quantum mechanics somehow arrives at the correct individual event, even though it is not obvious from the mathematical framework, how it does that. This led to the assumption of hidden variables underlying its mathematical formulation.
\begin{figure}
\centering
\includegraphics[width=\columnwidth]{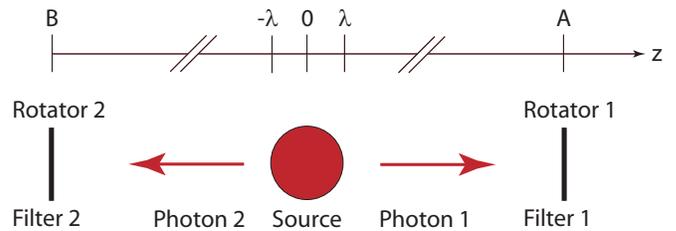}
\caption{{\bf EPR experiment.} Two photons are emitted from a common source, and subjected at two points $A$ and $B$ to separate polarization measurements, here a rotation of the fields and a filter.}\label{epr-fig}
\end{figure}

The solution here lies in the application of geometric algebra to rotations in three dimensional space and the insight that it is unnecessary to describe individual events as long as correlations are accurately described. This is shown by developing a new model for the experiments, which is as accurate but much more transparent than the conventional model in quantum mechanics. For all practical purposes photons can be assumed to be described by a variable Poynting-like vector ${\bf S}({\bf r},t) = {\bf e}_3 S({\bf r},t)$ parallel or antiparallel to the direction of motion. A multiplication of such a vector with a geometric product ${\bf e}_1 {\bf e}_2$ creates imaginary numbers \cite{doran2002}:
\begin{equation}
\left({\bf e}_1 {\bf e}_2\right) {\bf e}_3 S({\bf r},t) = {\bf e}_1 {\bf e}_2 {\bf e}_3 S({\bf r},t) = i S({\bf r},t)
\end{equation}
However, a geometric product is also a rotator; it thus rotates the photon's fields in the $xy$-plane perpendicular to the direction of motion. This fact can be used to develop a simple model of an EPR experiment with two photons emitted from a common source and measured by polarizers at points along the $z$-axis $A$ and $-B$, respectively.
If we rotate the original vector in the exponent, then a rotation affects the fields of the photons in the following way ($0 \le z_i \le \lambda, \varphi_i = z_i\,2 \pi/\lambda$):
\begin{eqnarray}
R(A) &=& e^{\left({\bf e}_1{\bf e}_2\right) {\bf e}_3 z_1\,2 \pi/\lambda} = e^{i \varphi_1} \nonumber \\
R(B) &=& e^{-\left({\bf e}_1{\bf e}_2\right) {\bf e}_3 z_2\,2 \pi/\lambda} = e^{-i \varphi_2}
\end{eqnarray}
Including the phase difference at the source $\Delta$ and analyzing the effect of two separate rotations at either end it is seen immediately that the two rotations are in fact connected by a complex phase:
\begin{equation}
R(A) \cdot R(B) = e^{i(\varphi_1 - \varphi_2 - \Delta)}
\end{equation}
This means that even though the two rotations are independent, they cannot be described by a product of two real variables. In our view, this was the key mistake of John Bell, which makes the Bell inequalities inapplicable to such a situation \cite{bell64}. The final step in the model is to consider probabilities of detecting a single photon after one rotation, and the joint probability of detecting both photons after two rotations. Here, one must introduce an initial phase $\varphi_0$, because the initial phase of a single photon is unknown. If the initial phase is equally probable for every value within the interval [$0,2\pi$], then a single measurement related to the final phase after rotation will be independent of the rotation angle. This is the result of the experiments. In our model it is captured by a probability of detection given by:
\begin{eqnarray}
p(A) &=& [\Re (R(A) \cdot e^{i \varphi_0})]^2 = \cos^2\left(\varphi_1 + \varphi_0\right)\\
p(B) &=& [\Re (R(B) \cdot e^{-i \varphi_0})]^2 = \cos^2\left(\varphi_2 + \Delta + \varphi_0\right) \nonumber
\end{eqnarray}
However, while single measurements will be statistically distributed due to the initial phase, coincidence measurements will not, as the initial phase cancels out. For $\Delta = 0$ we get:
\begin{equation}
p(A,B) = p\left[\Re (R(A) \cdot R(B))\right]^2 = \cos^2 \left(\varphi_1 - \varphi_2 \right)
\end{equation}
Using the Clauser-Horne-Shimony-Holt inequalities \cite{CHSH}, and setting coincidence counts equal to the coincidence probabilities in our model:
\begin{eqnarray}
C^{++} = C^{--} &=& \cos^2\left(\varphi_1 - \varphi_2\right) \nonumber \\
C^{+-} = C^{-+} &=& 1 - \cos^2\left(\varphi_1 - \varphi_2\right)
\end{eqnarray}
one arrives at the standard expectation values for coincidences \cite{hofer2012b}:
\begin{equation}
E(\varphi_1,\varphi_2) = 2 \cos^2\left(\varphi_1 - \varphi_2\right) - 1 = \cos \left[2\left(\varphi_1 - \varphi_2\right)\right]
\end{equation}
The standard result in quantum mechanics, which violates the Bell inequalities for a sum $S$ is recovered, if a certain set of angles is chosen for the rotations.
\begin{eqnarray}
S\left(\varphi_1,\varphi_1',\varphi_2,\varphi_2'\right) &=& E(\varphi_1,\varphi_2) - E(\varphi_1,\varphi_2') \\ &+&
E(\varphi_1',\varphi_2) + E(\varphi_1',\varphi_2') = 2 \sqrt{2} \nonumber
\end{eqnarray}
if  $\varphi_1 = 0, \varphi_1' = 45, \varphi_2  = 22.5, \varphi_2' = 67.5$. Note that the model is still statistical, and it is without doubt a local model. This seems to relate to Khrennikov's analysis, that non-locality is not a necessary condition for a violation of the Bell inequalities \cite{khrennikov08}. In addition, the only hidden variable in this description is a complex phase, which is the result of rotations in geometric algebra. Since this model reproduces all results in the standard model in quantum mechanics, it is hard to see how the standard theory could actually contain hidden variables in the form generally assumed.

Now let us assume that the difference between the two angles is a multiple of $\pi/2$. In this case the probabilities will be:
\begin{equation}
p(\varphi_1 - \varphi_2 = \pi/2) = 0 \qquad p(\varphi_1 - \varphi_2 = \pi) = 1
\end{equation}
Consider now a change of angle $\varphi_1$ during the experimental run, so that in one case, $\varphi_{1,0}$ the difference between the two angles is $\pi/2$, in the other case, $\varphi_{1,1}$ the difference is $\pi$. It is still impossible to predict the result at $A$, since this will depend, as the single polarization measurement, on the initial phase. However, it is possible to predict the outcome at $B$ if $A$ is known. The measured results at $A$ and $B$ will be:
\begin{eqnarray}
\begin{array}{cccccccc}
\varphi_{1,0}: & A = + & \Rightarrow & B = - & \varphi_{1,1}: & A = + & \Rightarrow & B = + \\
               & A = - & \Rightarrow & B = + &      & A = - & \Rightarrow & B = -\\
\end{array}\nonumber
\\
\end{eqnarray}
Given that the experimental outcome of $B$ depends on both, the setting of angle $\varphi_1$ and the outcome at $A$, it is claimed that the experiments are the consequence of
some {\it spooky action at a distance} \cite{EPR}. Since the angle can be changed in-flight, also the experimental result of $B$ given a result $A$ seems to change in this time-span. However, the result at $A$ is not known initially, neither is the result at $B$. Therefore the change of the angle $\varphi_1$ does not determine the outcome at $B$, it only affects the correlation. As this analysis shows, the effect is neither: neither spooky, nor action at a distance. That the measurements were at all interpreted in these terms is a classical error in logical analysis, which is captured by the phrase taught in most undergraduate physics courses: {\em correlation is not causation}.

\subsection{Single and double-slit interferometry}

Feynman called this experiment '' a phenomenon which is impossible, absolutely impossible, to explain in any classical way, and which has in it the heart of quantum mechanics. In reality, it contains the only mystery'' \cite{feynman-lectures}. Interestingly, all features of atomic scale interference experiments can be reproduced experimentally on the millimeter or centimeter scale, using the ingenious method developed by Y. Couder's group \cite{couder06}. The key element of the arrangement in Couder's experiments is an oscillating surface with a particular frequency of oscillation, along which a single droplet diffuses. Physically, this can be seen as a discrete set of interactions of the droplet with the surface, which will lead to a discrete set of momentum changes in horizontal direction. Interactions with a solid would also contain such a discrete set of interactions, as the atoms are in a regular arrangement. The wave properties are in this case not intrinsic to an electron or photon, they are externally imposed due to the interactions with the slit system. This point was well illustrated in a talk by Gerhard Gr\"ossing \cite{groessing10}, who developed a model of interference for single particles based on a quantized zero-point field directing the motion of individual particles. There is thus no mystery, and no intrinsic wave property involved, just a discrete set of interactions with the measuring device. Experimentally,
double-slit interferometry experiments of photons have already revealed that (i) a single photon will pass through a single slit, and (ii) the intensity at a particular point of the detector screen is primarily composed of photons passing through the adjacent slit \cite{steinberg2011}. The only viable interpretation of the experiments in this case is that the {\em interaction with the slit system itself}, and not a virtual extension of individual photons creates the maxima and minima of the interference pattern. A model of interference of electrons to this effect is in preparation.

\section{Discussion}

The emerging picture, from the preceding sections, is one of a scientific revolution with a depth and scale not seen since the quantum revolution itself, about a century ago.
However, on closer scrutiny, the key elements of our existing understanding of atomic scale physics remain in place. The Schr\"odinger equation still describes the properties and dynamics of electrons, albeit with wavefunctions, which no longer can be seen as primary entities. Condensed matter physics, which is to a large extent described by an adaptation of the Schr\"odinger equation in density functional theory, also remains largely the same, even though the detailed aspects of magnetic properties and the local characteristics of spins and their dynamics may be subject to revision. Given that condensed matter physics is also the basis for our understanding of chemistry and biology, all these sciences remain to a large extent unchanged in their basic understanding.

What is removed, is the additional weight quantum mechanics carried with it in the form of contradictions, paradoxes, impossibilities, and plain weirdness. There is no quantum weirdness left, once the extension of electrons, the role of wavefunctions, the specifics of rotations in three dimensional space, and the consequences of discrete interaction energies and momenta are thoroughly understood. This will almost certainly not be welcomed by some colleagues: after all, this quantum weirdness made for hugely exciting research programs and research papers for the last two generations. It remains to be seen, which of the more outlandish predictions, possible only within the ill-defined conceptual framework of conventional quantum mechanics, will in the end survive.

\section{Outlook}

While on the face of it not too much seems to change in physics, chemistry, or biology at the level of atoms and molecules, the situation could be completely different at the very small scale and at the very large scale. If electrons are extended, then
the emission of one electron by a single neutron has to be understood somewhat differently than is currently the case in the so-called standard model. In this model the electron is assumed to be created during the decay process of the neutron. However, extended electrons would be part of the neutron before emission, and change their density once they are free from the central proton. Such a model thus involves not the creation, but the phase change of the involved electron. In this case a high-density phase of electrons would also be part of every atomic nucleus containing neutrons. The conceptual framework for nuclear physics will then completely change and possibly resemble a lattice model in condensed matter physics more than the current models with their assembly of nuclear particles and interactions. How such a model can be developed remains to be seen.

On the very large scale, the situation would also be completely different than it is now. One of the key features of this new model of electrons and its embedding into atomic physics is the removal of the usual divide between the quantum world and the classical world. The wavefunctions in three dimensional space, formulated as multivectors, will be continuous for every continuous material. There is thus no difference between the model of a solid body in macro physics, and the model of the same body in microphysics: in both cases the wavefunctions and the wave-like structure of the electron's density will constitute the electron charge distribution throughout the solid body with a bivector potential capturing interactions of electrons in the body. While in today's many-body theories wavefunctions are defined in phase space and their information content increases exponentially as the size of the system increases, the time-independent wavefunctions of extended electrons, even for a system of many electrons, only has four basic variables: the square root of electron density and spin,
$\rho^{1/2}$ and $S^{1/2}$, and the direction of the unit vector ${\bf e}_S$. In this case wavefunctions are not limited by the van Vleck catastrophe and sensible scientific concepts even for large systems. Mathematically, one could use coarse-grained methods to remove some of the numerical effort to solve the Schr\"odinger equation of the whole body, but one would still have to account for the approximate nature of such a description by conceding the possible errors in the result of a prediction. However, gravitational forces on this level are of course part of the whole description of the system, albeit forces with a much smaller intensity, but not, in principle, a different nature. There is then no divide between gravity, electromagnetism, and electron dynamics. How this will affect our understanding of mesoscopic and macroscopic systems remains to be seen.


\section*{Acknowledgments}
I am grateful to the organizers, in particular Andrei Khrennikov, for giving me the opportunity to take part in this conference, which seems to have moved the field of quantum mechanics to a much deeper level of scrutiny \cite{khrennikov12}.






\end{document}